# Nuclear Spin Crossover in Dense Molecular Hydrogen

Thomas Meier[1*], Dominique Laniel[2], Miriam Pena-Alvarez[3], Florian Trybel[1], Saiana Khandarkhaeva[1], Alena Krupp[1], Jeroen Jacobs[4], Natalia Dubrovinskaia[2], Leonid Dubrovinsky[1]

1) Bayerisches Geoinstitut, University of Bayreuth, Bayreuth, Germany
2) Material Physics and Technology at Extreme Conditions, Laboratory of Crystallography, University of Bayreuth, Bayreuth, Germany
3) Centre for Science at Extreme Conditions and School of Physics and Astronomy, University of Edinburgh, Edinburgh, United Kingdom
4) European Synchrotron Radiation Facility (ESRF), Grenoble Cedex, France

**The laws of quantum mechanics are often tested against the behaviour of the lightest element in the periodic table, hydrogen. One of the most striking properties of molecular hydrogen is the coupling between molecular rotational properties and nuclear spin orientations, giving rise to the spin isomers ortho- and para-hydrogen. At high pressure, as intermolecular interactions increase significantly, the free rotation of $H_2$ molecules is increasingly hindered, and consequently a modification of the coupling between molecular rotational properties and the nuclear spin system can be anticipated. To date, high-pressure experimental methods have not been able to observe nuclear spin states at pressures approaching 100 GPa[1,2] and consequently the effect of high pressure on the nuclear spin statistics could not be directly measured. Here, we present *in-situ* high-pressure nuclear magnetic resonance data on molecular hydrogen in its hexagonal phase I up to 123 GPa at room temperature. While our measurements confirm the presence of I=1 ortho-hydrogen at low pressures, above 70 GPa, where inter- and intramolecular distances become comparable, we observe a crossover in the nuclear spin statistics from a spin-1 quadrupolar to a spin-1/2 dipolar system, evidencing the loss of spin isomer distinction. These observations represent a unique case of a nuclear spin crossover phenomenon in quantum solids.**

Changes in electronic spin statistics under changing thermodynamic conditions are an established physical crossover phenomenon[3]. It has direct applications for spintronics[4] and enables the understanding of the stabilization of magnetospheres of rocky (Earth-like) planets[5] as well as gas- and ice-giants[6]. The degrees of freedom of the nuclei spins, however, are widely approximated as fixed within the analysis of experiments, due to large shielding by core electrons and the extremely short atomic distances necessary to induce such fundamental changes.

Hydrogen, on the other hand, exhibits no core electrons and when bound contributes its electron to the molecular bond. Furthermore, the thermal de-Broglie wavelength of a hydrogen atom ($\lambda_{th}^H$) – the length scale under which atomic interactions require quantum-mechanical treatment – is about 1.1 Å at room temperature; significantly larger than for all other known elements. The combination of both effects results in a number of fascinating physical phenomena in molecular $H_2$ [7–9].

One property intrigued physicists in particular: the nature of the nuclear spin of the $H_2$ molecule and the resulting coexistence of the spin isomers ortho- (ortho-$H_2$) and para-hydrogen (para-$H_2$). Following Pauli`s exclusion principle, in order for the total $H_2$ molecular wave function to be antisymmetric under exchange of atomic positions, demands for the rotational ground state J=0, that the corresponding total nuclear wave function is antisymmetric (singlet state of I=0, i.e. para-$H_2$). Analogously, for the J=1 rotational state, the total nuclear wave function is required to be symmetric (triplet state of I=1, i.e. ortho-$H_2$). Therefore, the spin allotropic isomerism of the $H_2$ molecule originates in the coupling of both rotational state and nuclear spin. It has been argued[10,11] that at high pressure (*P*) ortho- and para-hydrogen spin isomers remain stable up to the dissociative Wigner-Huntington transition at $P > 400$ GPa[12,13]. This can only be assumed for weak or moderate intermolecular interactions, i.e. when nearest neighbour distances ($r_n$) are much shorter ($\approx 0.7$ Å at ambient conditions) than next-nearest neighbour distances ($r_{nn} \approx 3$ Å at ambient conditions), allowing for sufficient intramolecular wave function overlap (left side of Fig. 1a). Under high enough densities, however, intermolecular interactions increase significantly as $r_{nn}$ decreases by ~70% within 100 GPa[14,15]. At these pressures, $r_{nn}$ becomes comparable to $\lambda_{th}^H$ and collective nuclear quantum fluctuations increase rapidly[16]. For decades, theoretical[8,17,18] and experimental[9] studies have indicated that under such extreme pressures, the rotational quantum number J becomes less well defined, leading to a potential indistinguishability of the hydrogen spin isomers.

The only experimental technique in high-pressure research to directly study the $H_2$ nuclear spin states is nuclear magnetic resonance (NMR) spectroscopy, detecting the linear response of the nuclear spin system upon radio frequency excitation in a magnetic field $B_0$. The spin singlet state of para-hydrogen is NMR silent, whereas application of $B_0$ lifts the three-fold degeneracy of the ortho states and allows for an excitation of nuclear spin transitions (Fig. 1b). Nuclear spin pairing in ortho-hydrogen leads, furthermore, to a finite electric quadrupole moment, *eQ*, interacting with the local charge distribution defined by the structural arrangement of hydrogen molecules. Thus, quadrupolar coupling can be considered the dominant spin interaction, resulting in characteristic NMR line shapes[19] (Fig. 1c).

Here, we present $^1$H-NMR data of dense molecular hydrogen up to 123 GPa at room temperature. Details on experimental conditions, spectral simulations as well as data analysis are provided in the Methods and Materials Section.

Two NMR-DACs equipped with diamond anvils of 250 µm and 100 µm culets were loaded with molecular $H_2$. At low pressure, intense $^1$H resonances of roughly 500 kHz width were detected. With increasing *P*, the resonance signals broadened significantly approaching 750 kHz at 68 GPa (Fig. 2). Above 68 GPa, we observed a resonance narrowing accompanied by the emergence of two Pake doublets[20] with increasing splitting upon compression.

For the quadrupolar nature expected for ortho-$H_2$ (I=1), the electric field *V*(*r*), defined by the local charge distribution based on the crystal structure of phase I, should influence the shape of the observed resonance lines. Calculated line shapes for a I=1 spin system are shown in Figure 2 at pressures of up to 68 GPa. The order of magnitude of the quadrupolar interaction energy was considered small relative to the nuclear Zeeman energy[19] and consequently treated as a first order perturbation (see Methods and Materials Section for computational details). Up to 68 GPa, the measured $^1$H-NMR spectra are well described by these calculated line shapes broadened by first order quadrupole interaction. The line shape is mainly controlled by two parameters: (i) the quadrupole coupling constant $C_q$ describing the coupling between *eQ* and *V*(*r*) as well as (ii) the electric field gradient asymmetry parameter $\eta$ accounting for the geometry of *V*(*r*).

Figure 3a (top panel) shows estimated values of $C_q$ which increase from 28.1 kHz at 20 GPa to 61.9 kHz at 58 GPa. This increase is likely originated in the high compressibility and rapidly reducing next-nearest neighbour distances between molecular $H_2$ units, enhancing quadrupolar coupling. The asymmetry parameter $\eta$ (Fig. 3a, bottom panel) was found to be almost constant within experimental errors varying between 0.44 at 20 GPa and 0.49 at 58 GPa. Based on the hexagonal crystal structure of phase I[14,21], $\eta$ can be expected to be close to 0.5, which is in excellent agreement with values derived from the analysis of our NMR measurements.

Above $P \approx 68$ GPa, however, we observed a sudden decrease in both $C_q$ and $\eta$ coinciding with a resonance peak splitting. Since no structural rearrangement of $H_2$ molecules is reported by diffraction methods[21] or Raman spectroscopy[22] at room temperature in this *P*-range, effects based on a modification of the $H_2$ nuclear spin system should be considered.

This pressure indicates a turning point in the behaviour of the $^1$H spin system, as the observed peak splitting devolves from having dominantly quadrupolar characteristics towards a system controlled by nuclear dipole-dipole coupling, resulting in pronounced I=1/2 line shapes[23] with a frequency difference between spectral density function singularities directly correlated to the distances between hydrogen atoms. Considering that in this case both, the interaction with the nearest and next nearest neighbours will result in a dipolar NMR pattern, respectively, a superposition of signals as shown in Figure 1c can be expected. Computationally obtained values of nearest and next-nearest neighbour distances[15] are $r_\text{n} \approx 0.731$ Å and $r_\text{nn} \approx 1.342$ Å at 120 GPa. Calculating the distances from the parameter-set obtained through analysis of the NMR spectra for such a mixed scenario resulted in: $r_\text{n} \approx 0.727$ Å and $r_\text{nn} \approx 1.27(8)$ Å at 123 GPa, in excellent agreement with the computational estimates by Labet et al.[15]. Figure 2 shows the comparison between experimental spectra and calculated I=1/2 line shapes between 71 GPa and 123 GPa. The respective values derived for both nearest (top panel) and next-nearest (middle panel) neighbour distances can be found in Figure 3b. Additionally, the comparison between the equation of state derived from *ab-initio* computations [24] and diffraction data[14,21], along with the unit cell volumes (blue dots) derived from $r_\text{n}$ and $r_\text{nn}$ extracted from the analysis of the NMR spectra are shown in the bottom panel of Figure 3b.

Homonuclear Lee-Goldburg decoupling sequences[25] have been used to suppress quadrupolar and dipolar line broadening in order to resolve isotropic chemical shifts, $\delta_\text{iso}$. Figure 3c shows the evolution of $\delta_\text{iso}$: initially decreasing from 8.6 to 5.9 ppm between 20 and 59 GPa, $\delta_\text{iso}$ has an inflection point at ~60 GPa and raises under further compression to 22.7 ppm at 123 GPa. Comparison with Raman data[26] suggests that the minimum in $\delta_\text{iso}$ coincides with the well-known turn-over in the Raman shift of the $H_2$ vibron caused by a weakening of intramolecular and increased intermolecular interactions[27].

The presented data analysis leads to the following interpretation of the observed effects: At $P < 60$ GPa, $^1$H-NMR data is characteristic for a I=1 quadrupolar spin system as expected for ortho-$H_2$. In this regime, individual nuclear spin angular momenta couple with their nearest neighbours separated on average by $r_\text{n}$ ($< \lambda_\text{th}^H$) leading to a significant wave function overlap within the molecular units and a stabilisation of the spin isomers. The excellent agreement between NMR derived values for the electric field gradient asymmetry parameter $\eta$ and values inferred from X-ray diffraction data[14,21] strengthens this assessment. Following the theoretical study of Strzhemechny et al.[28], this pressure driven enhancement of the quadrupolar coupling constant $C_q$ in this *P*-regime may be interpreted as an experimental evidence for the mechanism of ortho-para conversion through electric quadrupole interaction.

At $P > 60$ GPa, quadrupolar coupling rapidly diminishes despite the absence of a structural rearrangement of the molecular $H_2$ units. Starting from about 70 GPa, spectral features characteristic of homonuclear dipole-dipole coupling between nearest and next-nearest neighbours become apparent. Provided the good agreement between NMR data with DFT [15] and experimentally[14,21] derived intermolecular and interatomic distances, this shift in behaviour implies intramolecular coupling of nuclear spins to become increasingly perturbed. The inflection-point in the isotropic chemical shift $\delta_{\text{iso}}$ strengthens this hypothesis as the increasing nuclear de-shielding above 60 GPa indicates a shift of electron density away from individual molecular centres towards intermolecular regions.

$^{14}$N-NMR on molecular nitrogen at P = 3 GPa (see Methods and Materials Section) supports this argument, since $\lambda_{\text{th}}^N \approx 0.3$ Å is smaller than $r_n \approx 1.2$ Å, the nitrogen spin system shows clear characteristics of a nuclear spin triplet state anticipated within the non-pairing regime contrary to the quintuplet state stabilised in the molecular spin pairing regime.

In conclusion, even at moderately high pressures (< 100 GPa) intramolecular nuclear spin coupling breaks down and the hydrogen spin system adopts an average dipolar I=1/2 value. Crossovers of the nuclear spin statistics of a quantum solid such as hydrogen have never been observed before and given the large compressibility of hydrogen in conjunction with a large $\lambda_{\text{th}}^H$, this crossover phenomenon might only be experimentally observable in molecular H$_2$. Nuclear spin statistics of similar diatomic molecules (e.g. N$_2$) are likely to be best described as non-pairing nuclear spins due to enhanced atomic masses -- and thus short $\lambda_{\text{th}}$ -- as well as reduced compressibilities due to the presence of core electrons.

The nuclear spin crossover may have far reaching consequences for understanding different phenomena, even seemingly unlinked to this study. Explanation of the stabilization of magnetospheres of gas- and ice-giants may serve as an example. Recently[6] it was argued that either the strong magnetic fields of Jupiter, Uranus and Neptune are a consequence of spin isomer driven hyperpolarization in their upper hydrogen-rich troposphere or originates from metallic hydrogen reservoirs in their inner cores. Our observations show that above 70 GPa at 300 K spin allotrope distinction vanishes, effectively terminating such hyperpolarisation processes. Furthermore, as a consequence of the inverse square root temperature dependence of $\lambda_{\text{th}}^H$, the crossover likely shifts towards lower pressures with decreasing temperatures, further reducing the likelihood of planetary magnetic field stabilization in shallow tropospheric depths of Jovian planets due to hydrogen spin isomer hyperpolarization.

**Acknowledgements**
We thank Nobuyoshi Miyajima for help with the FIB milling. We are very thankful to Graeme Ackland and Gerd Steinle-Neumann for fruitful discussions.

**Funding**

The authors thank the German Research Foundation (Deutsche Forschungsgemeinschaft, DFG, Project Nos. DU 954/11-1, DU 393/13-1, DU 393/9-2, STE 1105/13-1 and ME 5206/3-1) and the Federal Ministry of Education and Research, Germany (BMBF, Grant No. 05K19WC1) for financial support. D.L. thanks the Alexander von Humboldt Foundation for financial support. N.D. thanks the Swedish Government Strategic Research Area in Materials Science on Functional Materials at Linköping University (Faculty Grant SFO-Mat-LiU No. 2009 00971).

**Author Contributions**

T.M. and L.D. designed the experiment. T.M., S.K., A.K. and J.J. prepared the DACs and NMR resonators. T.M., D.L., M.P.A., F.T. and A.K. performed and analysed the experiments. T.M., M.P.A., D.L., F.T., N.D. and L.D. analysed the results and wrote the manuscript.

**Competing Interests**

The authors declare that they have no competing interests.

**Data availability**

The data supporting the findings of this study are available from the corresponding author upon reasonable request.

**Materials and Methods**

*Diamond Anvil Cell Preparation*

Two diamond anvil cells, equipped with pairs of diamond anvils with a culets size of 250 µm and 100 µm, were prepared. Rhenium gaskets were pre-indented to 25 µm and 10 µm respectively, and 80 µm and 40 µm diameter holes were laser drilled in the centre of the indentation to form the sample cavities, resulting in sample volumes of about 125 pl and 13 pl respectively.

The diamond anvils were coated with a 1 µm thick layer of copper using physical vapour deposition[1]. Double[2] (in the case of the 250 µm diamonds) and triple[3] (for the 100 µm diamonds) stage Lenz-lens radio-frequency resonators were produced by using focused ion beam milling. To ensure electrical insulation and avoid hydrogen diffusion into the rhenium, the gaskets were coated by 500 nm thick layers of $Al_2O_3$. Radio-frequency excitation coils were made from 100 µm thick, teflon insulated, copper wire and arranged such that a Helmholtz coil pair is formed.

Hydrogen loading was conducted at the ESRF at and pressure was increased at cryogenic temperatures to avoid rapid hydrogen diffusion into the diamond anvils. Pressure was calibrated by means of the diamond edge Raman scale[4,5]. Comparison of the vibron frequencies of the $H_2$ samples at elevated pressures shows a slight systematic offset of less than 5 GPa at the highest pressures where Raman data was collected[6].

*NMR Experiments*

All NMR experiments were conducted using a solid-state NMR spectrometer from *Tecmag Inc.* (Redstone) equipped with a 100 W pulse amplifier. To polarize the nuclear spin system, we used a sweepable electromagnet with an average magnetic field of 1 T and sufficiently high homogeneity. Intense $^1$H-NMR signals were recorded at frequencies of 45.26 MHz, corresponding to an external magnetic field strength of about 1063 mT. Using nutation experiments, we found optimal excitation pulses between 1 - 1.2 µs for both cells, in reasonable agreement with earlier experiments[1-3, 7].

Free induction decays were excited using a single pulse of 833 kHz to 1 MHz bandwidth. The spectrometer was blanked off for 2 µs to avoid damage to the pre-amplifier. Supplementary Figures S2 and S3 show all $^1$H-NMR spectra recorded by this method.

In order to resolve isotropic chemical shifts, $\delta_{iso}$, a Lee-Goldburg pulse for homonuclear decoupling was used as described earlier[7]. The resulting narrowed NMR spectra had a FWHM line width of about 3 ppm, thus the resolution accuracy of $\delta_{iso}$ was found to be in the order of 0.1 ppm (Figure S4). Resonance frequencies were referenced towards an aqueous solution of tetramethylsilane in a similar DAC at ambient pressure conditions.

*Computation of NMR Lineshapes and AsymmetryPparameters of the Electric Field Gradient*

Calculation of the NMR line shapes was carried out following the analytical method outlined by Bloembergen and Rowland[9], Pake[10] and Hughes and Harris[11]:

Using the standard expressions for the resonance frequency distribution ω for both first order quadrupole interaction as well as homonuclear dipole-dipole interaction:

$$\omega(\alpha, \beta, m) = \omega_Q \cdot (m + 1/2) \cdot \left(\frac{3\cos^2\beta - 1}{2} - \frac{\eta}{2} \sin^2\beta \cos(2\alpha)\right), \quad (1)$$

$$\omega_i(\alpha) = d_i \cdot \left(\frac{3\cos^2\beta - 1}{2}\right), \quad (2)$$

with

$$\omega_Q = \frac{6\pi}{2I(2I+1)} \cdot C_q, \quad (3)$$

$$C_q = \frac{e^2 qQ}{h}, \quad (4)$$

$$d_i = \frac{\mu_0 \gamma_n^2 \hbar}{8\pi^2 \, r_i^3}, \quad (5)$$

where the Euler angles α and β describe the orientation of the crystallites with respect to the external magnetic field. $\gamma_n$ is the gyromagnetic ratio of the hydrogen nuclei, m the nuclear spin quantum number (m = 1, 0, -1) and $r_i$ the average distance between interacting hydrogen nuclei, $r_n$ or $r_{nn}$, respectively. η describes the asymmetry of the electric field gradient tensor ($V_{ij}$) in the principal axis system as:

$$\eta = \frac{V_{yy}-V_{xx}}{V_{zz}}, |V_{zz}| > |V_{xx}| > |V_{yy}|. \tag{6}$$

The line shape function, $P(\omega)$, for quadrupolar spin interactions, is given by:

$$P(\omega) = \sum_m \int_{-1}^{1} \frac{\mu}{4\pi} \sin(\beta(\omega,\alpha,m)) \cdot \left(\left\|\frac{\partial \beta(\omega,\alpha,m)}{\partial \omega}\right\|\right) d(\cos(2\alpha)), \tag{7}$$

where β(ω, α, m) denotes the inverse function of eq. (1) with respect to β, and μ accounts for the multiplicity of spectral functions. For the dipolar interaction $P(\omega)$, is given by

$$P(\omega) = \sum_i \int_{-1}^{1} \frac{\mu}{4\pi} \sin(\beta(\omega_i,\alpha)) \cdot \left(\left\|\frac{\partial \beta(\omega_i,\alpha)}{\partial \omega_i}\right\|\right) d(\cos(2\alpha)), \tag{8}$$

where $\beta(\omega_i, \alpha)$ denotes the inverse function of eq. (2) with respect to β, and μ accounts for the multiplicity of spectral functions.

Cut-off frequencies of the resulting spectral line functions were chosen according to the Heaberlein convention for NMR shift tensors[12]. Spectral line broadening was accounted for by convolution of the total line shape function with a Voigtian line of defined Lorentzian and Gaussian widths. In order to fit the experimental data, the respective line shape function $P(\omega)$ is optimized by varying $C_q$ and η for quadrupolar coupling and $r_n$ and $r_{nn}$ for dipolar coupling. The corresponding Python scripts are available from the authors upon request. Table S1 summarises all fit parameters.

In order to calculate the asymmetry parameter η of the electric field gradient tensor in the spin-pairing regime, we used the second derivative of the electric potential, V(**r**), defined by the molecular center of gravity positions from diffraction measurements[13]:

$$V(\mathbf{r}) = \frac{e}{4\pi\epsilon_0} \sum_i \frac{1}{\sqrt{(x-x_i)^2 + (y-y_i)^2 + (z-z_i)^2}}, \tag{9}$$

$$V_{ij} = \frac{\partial V(\mathbf{r})}{\partial x_i \partial y_j}. \tag{10}$$

Using eq. (6) under consideration of the ordering of the components of $V_{ij}$ in the principal axis system allows computation of $\eta$ from crystallographic data.

Table S1: Fitting parameters of $^1$H-NMR spectra. $C_q$ is the quadrupole coupling constant, $\eta$ the asymmetry parameter of the electric field gradient tensor in the principle axis system, $r_n$ and $r_{nn}$ are the nearest and second nearest neighbour distances, respectively. The isotropic chemical shift, $\delta_{iso}$, was derived after homonuclear Lee-Goldburg decoupling.

| | 1st Order Quadrupole Interaction | | Dipole-Dipole Interaction | | Lee-Goldburg Decoupling $\delta_{iso}$ |
|---|---|---|---|---|---|
| $P$ in GPa | $C_q$ in kHz | $\eta$ | $r_n$ in Å | $r_{nn}$ in Å | in ppm |
| 20 | 28.1(6) | 0.44(6) | -- | -- | 8.665(112) |
| 24 | 27.9(8) | 0.43(4) | -- | -- | 7.363(112) |
| 36 | 30.0(7) | 0.50(7) | -- | -- | 6.429(112) |
| 42 | 32.7(5) | 0.52(4) | -- | -- | 5.951(112) |
| 47 | 35.2(6) | 0.44(6) | -- | -- | 5.928(125) |
| 50 | 44.8(4) | 0.59(3) | -- | -- | 5.905(114) |
| 54 | 48.0(3) | 0.46(7) | -- | -- | 5.924(150) |
| 58 | 61.9(7) | 0.49(8) | -- | -- | 6.139(120) |
| 68 | 43.9(9) | 0.37(9) | -- | -- | 6.670(173) |
| 71 | 24.5(8) | 0.20(9) | 0.736(5) | 1.509(14) | 7.280(127) |
| 77 | 18.6(7) | 0.15(7) | 0.733(5) | 1.457(13) | 8.479(195) |
| 85 | 16.6(3) | 0.14(6) | 0.732(5) | 1.430(11) | 10.612(149) |
| 97 | 15.0(4) | 0.10(5) | 0.732(5) | 1.340(12) | 15.351(100) |
| 106 | 20.9(6) | 0.04(1) | 0.731(5) | 1.307(14) | 18.895(153) |
| 115 | 19.2(9) | 0.05(7) | 0.729(5) | 1.270(11) | 21.323(147) |
| 123 | 19.5(9) | 0.02(7) | 0.727(5) | 1.270(18) | 22.673(154) |

## $^{14}$N-NMR of Molecular Mitrogen at 3 GPa

Molecular nitrogen was measured using natural isotopic composition, where the majority of molecules can be expected to be pairs of $^{14}$N nuclei. As $^{14}$N nuclei have a nuclear spin of I=1, one can expect a spin pairing scenario similar to molecular D$_2$: the para-N$_2$ states consist of a quintuplet subsystem with I=2 whereas the ortho-N$_2$ states are a triplet subsystem.

The electric field gradient asymmetry parameter η was estimated according to diffraction data[14] to be around 0.23. Recorded $^{14}$N-NMR spectra (Figure S2; right panel) do not show pronounced shoulder, expected for a I=2 quadrupolar powder pattern in absence of $m_{-2 \to -1}$ and $m_{1 \to 2}$ transitions. In fact, the spin system is well described by a I=1 spin system using the estimated value for η (Figure S2; left panel).

According to structural data[14], $r_\mathrm{n}$ can be estimated to be around 1.2 Å at this pressure; four times longer than the thermal de-Broglie wavelength of a single $^{14}$N atom. Therefore, the wave function overlap should be negligible in molecular nitrogen at these pressures and nuclear spins considered unpaired.

## Spin Pairing regime (I = 1)

$$r_{nn} \gg \lambda_{th}^H = \frac{h}{\sqrt{2\pi m_p k_B T}}$$

## Non-Pairing regime (I = ½)

$$r_{nn} \gtrsim \lambda_{th}^H \gtrsim r_n$$

a)
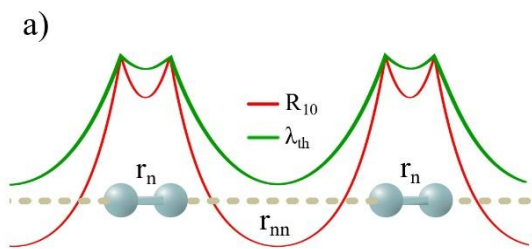
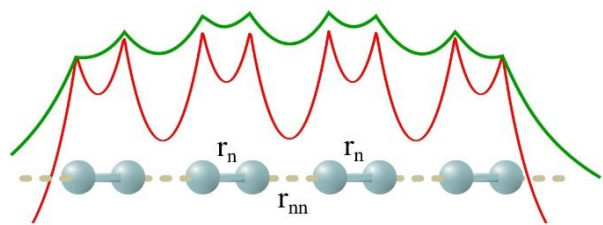

b)
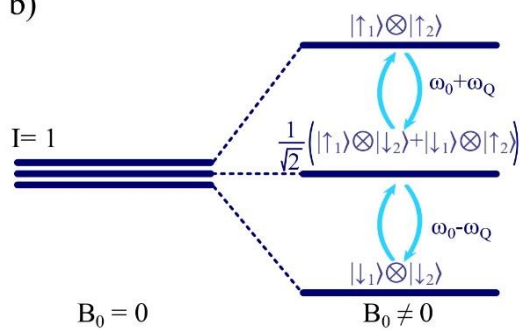

c)
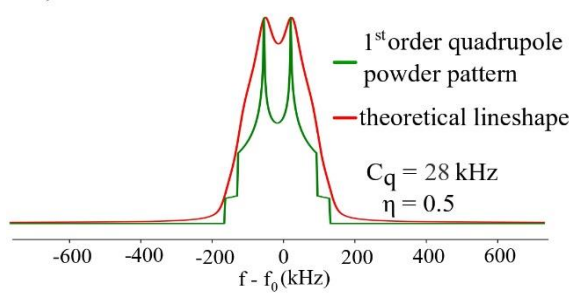
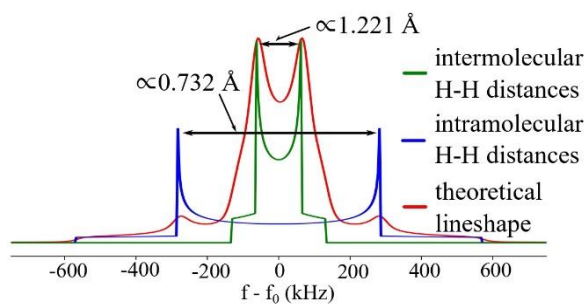

*Figure 1:* **Overview of both spin-pairing and non-pairing regimes. a)** Schematic representation of the wave function overlap (red) and the thermal de-Broglie wavelengths (green) of $H_2$ molecules. **b)** Schematic representation of the nuclear spin energy levels under the influence of an external magnetic field $B_0$ for the pairing (i.e. quadrupole interaction) and non-pairing (dipole-dipole interaction) regimes. **c)** Theoretical line shapes in the pairing and non-pairing regime.

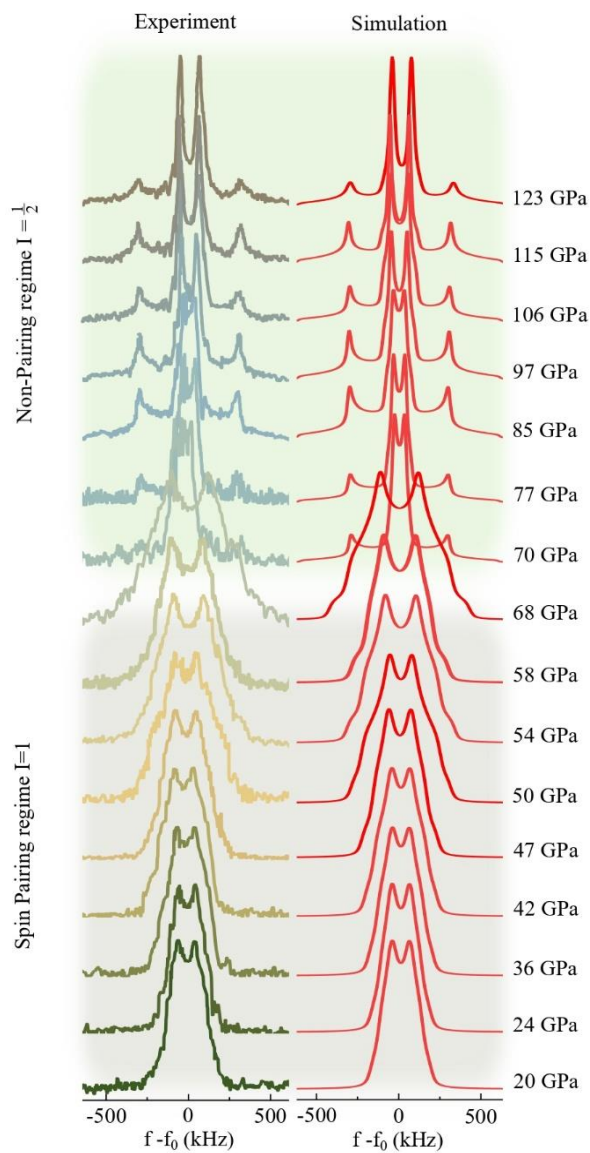

*Figure 2:  **Experimental and calculated ¹H-NMR spectra of molecular H₂ up to 123 GPa at room temperature.**.
Between 20 and 68 GPa, first order quadrupole interactions describe the experimental data reasonably well. At P > 68 GPa, spectra were found to be broadened by dipole-dipole interaction resulting in a superposition of two Pake doublets corresponding to nearest and next-nearest hydrogen distances.*

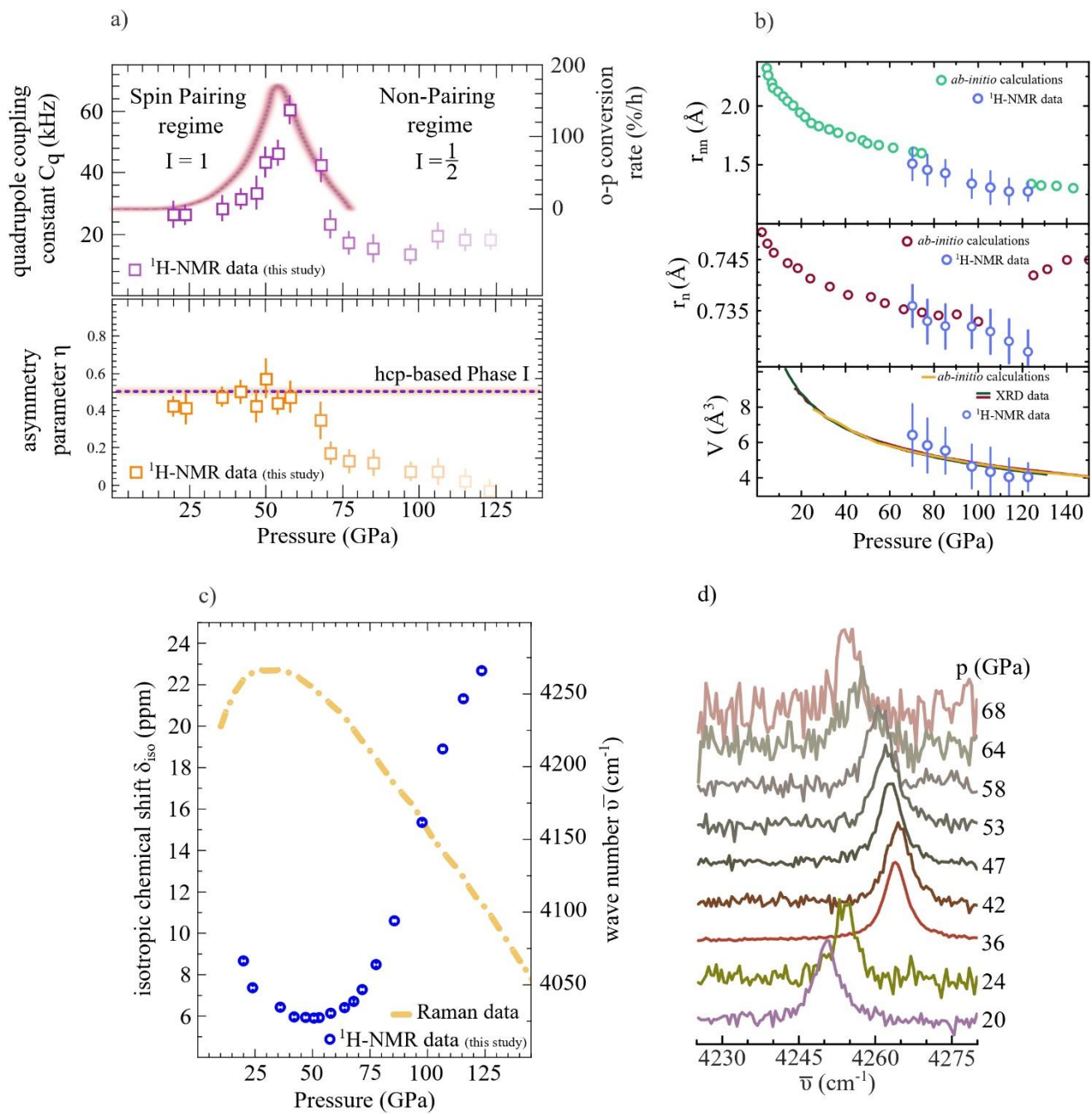

Figure 3: **Extracted ¹H-NMR data of molecular H₂ at pressures up to 123 GPa at room temperature. a)** Top panel: quadrupole coupling constant $C_q$ determined from NMR data (squares). The rose line denotes theoretical ortho-para conversion rates from electric quadrupole interaction[28]. Bottom panel: asymmetry parameter $\eta$ in the spin pairing regime (P < 60 GPa). The dashed line represents the inferred $\eta$ based on the hcp structure of Phase I from diffraction experiments. The shading of the squares at P > 60 GPa highlights the crossover to the non-pairing I=1/2 regime. **b)** Top panel: next nearest neighbour distances $r_{nn}$. Green circles are based on DFT computations[15]. Blue circles are extracted values of $r_n$ and $r_{nn}$ from the NMR spectra in the non-pairing I=1/2 regime. Middle panel: nearest neighbour distances $r_n$ (blue circles) and DFT calculations[15] (red circles). The discontinuity at P > 125 GPa in the DFT calculations indicates a transition from the hcp based to a monoclinic structure. Bottom panel: comparison between the equations of state derived from ab-initio computations[24] (yellow line) and diffraction data[14,21] along with the unit cell volumes (blue dots) derived from $r_{nn}$ and $r_n$ extracted from the NMR experiments. **c)** Extracted isotropic chemical shift values $\delta_{iso}$ after homo-nuclear Lee-Goldberg decoupling. Error bars are within the symbol size. The orange dotted line shows the room temperature Raman shift of the H₂ vibron[26]. **d)** Selected Raman spectra of the H₂ vibron **at increasing pressure.**

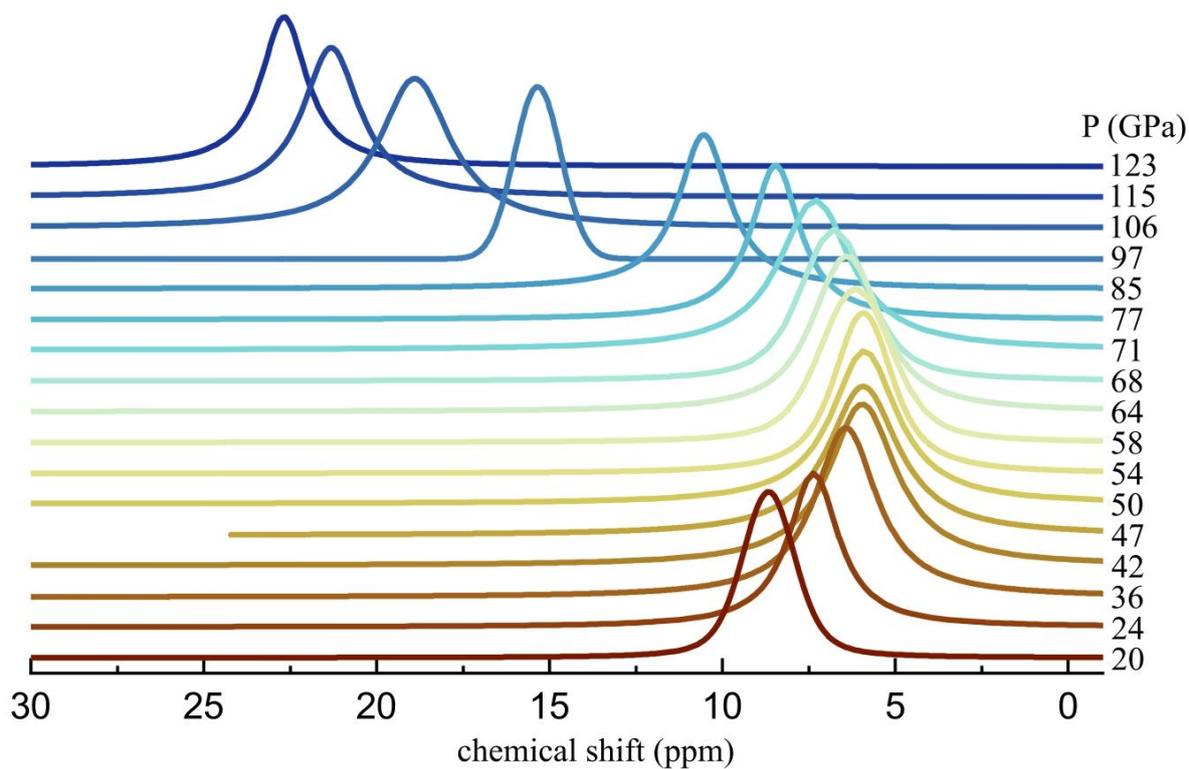

*Figure S1: **High Resolution ¹H-NMR spectra of molecular hydrogen**. Spectra were recorded using a Lee-Goldburg pulse scheme for homonuclear decoupling.*

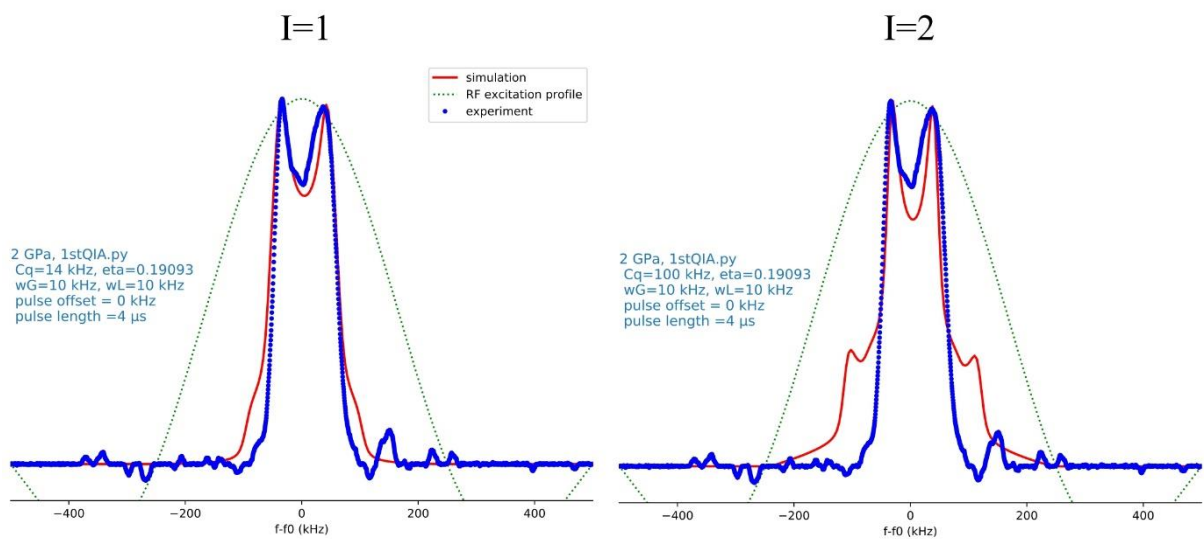

*Figure S2: **¹⁴N-NMR spectra of molecular nitrogen at pressures of about 3 to 4 GPa**. Left) simulation of a I=1 lineshape. Right) simulation of a I=2 lineshape.*